\documentclass[showpacs,twocolumn,preprintnumbers,amsmath,amssymb]{revtex4}

\usepackage{graphicx}
\usepackage{dcolumn}
\usepackage{bm}

\newcommand{\bq}{\begin{equation}}
\newcommand{\eq}{\end{equation}}
\newcommand{\bqa}{\begin{eqnarray}}
\newcommand{\eqa}{\end{eqnarray}}
\newcommand{\nn}{\nonumber \\}

\def\be     {\begin{equation}}
\def\ee     {\end{equation}}
\def\bea        {\begin{eqnarray}}
\def\eea        {\end{eqnarray}}
\def\bnn    {\begin{eqnarray*}}
\def\enn    {\end{eqnarray*}}

\begin{document}

\title{Distribution of critical temperature at Anderson localization}
\author{Rayda Gammag$^{1,2}$ and Ki-Seok Kim$^{3,4}$}
\affiliation{$^{1}$Asia Pacific Center for Theoretical Physics, POSTECH, Pohang, Gyeongbuk 790-784, Korea \\
$^{2}$Max Plank POSTECH Center for Complex Phase Materials, POSTECH, Pohang 790-784, Korea \\
$^{3}$Department of Physics, POSTECH, Pohang, Gyeongbuk 790-784, Korea \\
$^{4}$Institute of Edge of Theoretical Science, POSTECH, Pohang, Gyeongbuk 790-784, Korea }
\date{\today}

\begin{abstract}
Based on a local mean-field theory approach at Anderson localization, we find a distribution function of critical temperature from that of disorder. An essential point of this local mean-field theory approach is that the information of the wave-function multifractality is introduced. The distribution function of the Kondo temperature ($T_{K}$) shows a power-law tail in the limit of $T_{K} \rightarrow 0$ regardless of the Kondo coupling constant. We also find that the distribution function of the ferromagnetic transition temperature ($T_{c}$) gives a power-law behavior in the limit of $T_{c} \rightarrow 0$ when an interaction parameter for ferromagnetic instability lies below a critical value. However, the $T_{c}$ distribution function stops the power-law increasing behavior in the $T_{c} \rightarrow 0$ limit and vanishes beyond the critical interaction parameter inside the ferromagnetic phase. These results imply that the typical Kondo temperature given by a geometric average always vanishes due to finite density of the distribution function in the $T_{K} \rightarrow 0$ limit while the typical ferromagnetic transition temperature shows a phase transition at the critical interaction parameter. We propose that the typical transition temperature serves a criterion for quantum Griffiths phenomena vs. smeared transitions: Quantum Griffiths phenomena occur above the typical value of the critical temperature while smeared phase transitions result at low temperatures below the typical transition temperature. We speculate that the ferromagnetic transition at Anderson localization shows the evolution from quantum Griffiths phenomena to smeared transitions around the critical interaction parameter at low temperatures.
%
%
\end{abstract}


\maketitle

\section{Introduction}

Harris criterion \cite{Harris} deals with the stability of criticality against weak randomness in an average sense. Suppose a spin system with a critical temperature $T_{c}$, reduced by disorder. Since disorder breaks the translational symmetry, it is natural to consider an average local critical temperature $\langle T_{c}(\bm{x}) \rangle$ in a correlated volume $\xi^{d}$, where $\xi$ is the correlation length related with the criticality and $d$ is the spatial dimension. When variation of the average local critical temperature in the correlated volume, $\Delta \langle T_{c}(\bm{x}) \rangle \sim \xi^{- d / 2}$, is smaller than distance from the global ordering temperature, $t \sim \xi^{-1 / \nu}$ with the correlation-length critical exponent $\nu$, the Harris criterion of $\Delta \langle T_{c}(\bm{x}) \rangle < t$ thus $d \nu > 2$ tells us that the nature of the clean critical point is stable against weak randomness. When the Harris criterion is violated, i.e., $d \nu \leq 2$, disorder becomes relevant at the critical point, expected to change the nature of the clean critical point. If the resulting fixed-point value of disorder turns out to be finite, the Harris criterion is fulfilled with a modified correlation-length critical exponent $\nu'$ at such a disordered critical point. In particular, if the strength of randomness continues to increase toward infinity, referred to as an infinite randomness fixed point, the resulting disorder physics is governed by extreme inhomogeneity of the system \cite{Vojta_Review,Vlad_Review} and thus, the average sense is not much meaningful. In this situation local ``ordering" is allowed although macroscopic coherence is prohibited, the region of which is called rare region. ``Rare" in the name originates from an exponentially low probability to find such a region, given by $p_{rr}(L) \sim \exp ( - c L^{d} )$ with a positive numerical constant $c$, where $L$ is the length scale of the region and the exponent is an energy of the region. Such a rare region behaves as a super spin whose dynamics is extremely slow, being responsible for singularity in free energy and thus, dominating thermodynamics, referred to as Griffiths singularity \cite{Griffiths,McCoy_Wu}. As a result, rare region effects dominate critical physics not only at but also near the strong disorder critical point, called Griffiths phase.

Such rare region effects can be described by an averaged susceptibility $\chi_{av} \sim \int_{0}^{L_{sys}} d L p_{rr}(L) \chi_{rr}(L)$ from the rare region susceptibility $\chi_{rr}(L)$. Based on the rare region susceptibility, rare region effects have been classified into three categories \cite{Vojta_Review,Vlad_Review}. Class A: The rare region susceptibility is given by $\chi_{rr}(L) \sim L^{a}$. As a result, the Griffiths singularity is weak, essentially unobservable. Class B: The rare region susceptibility is given by $\chi_{rr}(L) \sim \exp(a L^{d})$ with a positive numerical constant $a$. As a result, the averaged susceptibility diverges inside the Griffiths phase. This happens when the rare region lies at the lower critical dimension, prohibiting the rare region from static ordering and promoting quantum tunneling between degenerate ordered states. Class C: When the rare region is above the lower critical dimension, the rare region susceptibility already shows divergence at its finite size $L$, meaning that such a rare region experiences a phase transition toward an ordered state at least locally and tunneling events are suppressed. As a result, they become randomly frozen and the phase transition is smeared \cite{Smeared}. Actually, this argument based on the lower critical dimension of the rare region serves a fascinating criterion in the case of strong disorder. However, it is not straightforward to have quantitative predictions within this criterion, in particular, at finite temperatures.

In this paper we revisit this long standing issue on the criterion of quantum Griffiths phenomena and smeared phase transitions. We study quantum phase transitions at the critical disorder strength of Anderson localization, physics of which are governed by the extreme inhomogeneity as discussed before. An idea is to find distribution of critical temperature from that of disorder. Focusing on rare regions, we develop a local mean-field theory to describe their ordering behaviors, where the wave-function multifractality \cite{Multifractality} is introduced into the self-consistent equation of an order parameter \cite{PTK1,PTK2,PTC}. As a result, we find a distribution function for critical temperature, where both the Stoner transition and the Kondo effect are examined. The distribution function of the Kondo temperature ($T_{K})$ shows a power-law divergent character in the $T_{K} \rightarrow 0$ limit regardless of the Kondo coupling constant. On the other hand, the distribution function of the ferromagnetic transition temperature ($T_{c}$) displays an abrupt change from a power-law divergent behavior to a vanishing tendency in the $T_{c} \rightarrow 0$ limit, increasing an interaction parameter for ferromagnetic instability across a critical value. The critical value turns out to be slightly larger than that of the clean limit. The typical transition temperature given by a geometric average reflects the power-law divergent character of the distribution function in the regime of low critical temperatures. Thus, these results imply that the typical Kondo temperature always vanish due to finite density of the distribution function while the typical ferromagnetic transition temperature shows a phase transition at the critical interaction parameter. This leads us to propose a criterion for quantum Griffiths phenomena and smeared phase transitions at finite temperatures: Above the typical transition temperature, quantum Griffiths phenomena occur while below it, smeared phase transitions result. We suggest that the ferromagnetic transition at Anderson localization shows the evolution from quantum Griffiths phenomena to smeared transitions around the critical interaction parameter at low temperatures. It should be pointed out that although the lower critical dimension of a rare region is replaced with a typical value of the critical temperature for local ordering, these two points of views are not inconsistent.

%
%

Two critical assumptions have been made: One is applicability of mean-field theory to a rare region of the nano-scale and the other is independence between rare regions. Both approximations can be improved by introducing fluctuation corrections into the mean-field description \cite{Fluctuation_Corrections} and taking into account couplings between rare regions, respectively.

\section{Local mean-field theory at the Anderson metal-insulator transition}

\subsection{Stoner instability}

\subsubsection{Formulation}

We start from an effective Hubbard Hamiltonian
\bqa
&& \mathcal{H} = \int d^{d} \boldsymbol{r} \Big\{ c_{\sigma}^{\dagger}(\tau,\boldsymbol{r}) \Bigl( - \frac{\boldsymbol{\nabla_{\boldsymbol{r}}}^{2}}{2m} - \mu + v(\boldsymbol{r}) \Bigr) c_{\sigma}(\tau,\boldsymbol{r}) \nn && + U c_{\uparrow}^{\dagger}(\tau,\boldsymbol{r}) c_{\uparrow}(\tau,\boldsymbol{r}) c_{\downarrow}^{\dagger}(\tau,\boldsymbol{r}) c_{\downarrow}(\tau,\boldsymbol{r}) \Big\}  ,
\eqa
where $c_{\sigma}(\tau,\boldsymbol{r})$ is an electron annihilation operator at time $\tau$ and position $\bm{r}$ with spin $\sigma$. $U$ is an effective local interaction parameter, $\mu$ is an electron chemical potential, and $v(\boldsymbol{r})$ is an external electric potential, randomly distributed.

In order to investigate ferromagnetic quantum phase transitions in disordered metals, we perform the Hubbard-Stratonovich transformation for the spin-triplet channel. Taking into account the disorder average in the presence of random electric potentials, we reach the following expression for an effective free energy
\bqa && \mathcal{F} = - \frac{1}{\beta} \int_{-\infty}^{\infty} d v(\boldsymbol{r}) P[v(\boldsymbol{r})] \ln \int D c_{\sigma}(\tau,\boldsymbol{r})
D \boldsymbol{\Phi}(\tau,\boldsymbol{r}) \nn && \exp \Bigl[ - \int_{0}^{\beta} d \tau \int d^{d} \boldsymbol{r}
\Bigl\{ c_{\sigma}^{\dagger}(\tau,\boldsymbol{r}) \Bigl( \partial_{\tau} - \mu + \frac{U}{6}
- \frac{\boldsymbol{\nabla_{\boldsymbol{r}}}^{2}}{2m} \nn && + v(\boldsymbol{r}) \Bigr) c_{\sigma}(\tau,\boldsymbol{r})
- c_{\sigma}^{\dagger}(\tau,\boldsymbol{r}) [ \boldsymbol{\Phi}(\tau,\boldsymbol{r})
\cdot \boldsymbol{\tau}]_{\sigma\sigma'} c_{\sigma'}(\tau,\boldsymbol{r}) \nn && + \frac{3}{2 U}
[\boldsymbol{\Phi}(\tau,\boldsymbol{r})]^{2} \Bigr\} \Bigr] , \label{Free_Energy_Start} \eqa
where $\boldsymbol{\Phi}(\tau,\boldsymbol{r})$ is an effective magnetic field, coupled to a spin-density field and determined self-consistently within the mean-field approximation. $P[v(\bm{r})]$ is a distribution function, given by the Gaussian $P[v(\bm{r})] = \mathcal{N}_{v} \exp\Bigl(- \int d^{d} \bm{r} \frac{v^{2}(\bm{r})}{2 \Gamma_{v}} \Bigr)$ for example, where $\Gamma_{v}$ is a variance and $\mathcal{N}_{v}$ is determined by the normalization condition $\int_{-\infty}^{\infty} d v(\bm{r}) P[v(\bm{r})] = 1$.

The basic idea is to reformulate this effective free energy, resorting to the eigenfunction basis for each configuration of random electric potentials, given by
\bqa && \Bigl( - \frac{\boldsymbol{\nabla_{\boldsymbol{r}}}^{2}}{2m} - \mu_{r}
+ v(\boldsymbol{r}) \Bigr) \Psi_{n}(\boldsymbol{r}) = \varepsilon_{n} \Psi_{n}(\boldsymbol{r}),
\label{Eigenfunction} \eqa
where $\Psi_{n}(\boldsymbol{r})$ is an eigenfunction with an eigenvalue $\varepsilon_{n}$ and an effective chemical potential $\mu_{r} = \mu - \frac{U}{6}$ in a fixed disorder configuration $v(\boldsymbol{r})$. Performing the ``Fourier transformation" of the electron field as
\bqa
&& c_{\sigma}(\tau,\bm{r}) = \frac{1}{\beta} \sum_{i\omega} e^{- i \omega \tau} \sum_{n} \Psi_{n}(\boldsymbol{r}) \psi_{\sigma n}(i\omega), \label{Fourier_Disorder}
\eqa
where $\psi_{\sigma n}(i\omega)$ is an electron field in the disorder basis, and rewriting the free energy [Eq. (\ref{Free_Energy_Start})] within this representation, we obtain an effective mean-field free energy
\bqa && \mathcal{F} = - \frac{1}{\beta} \int_{-\infty}^{\infty} d v(\bm{r}) P[v(\bm{r})] \ln \int D \psi_{\sigma n}(i\omega) \nn &&
\exp \Bigl[ - \sum_{i\omega} \sum_{n} \psi_{\sigma n}^{\dagger}(i\omega) \Bigl( [- i\omega + \varepsilon_{n}]
\delta_{nn'} \delta_{\sigma\sigma'} \nn && - \sum_{n'} \int d^{d} \bm{r} \Psi_{n}^{\dagger}(\bm{r}) \Psi_{n'}(\bm{r})
\bm{\Phi}(\bm{r}) \cdot \bm{\sigma}_{\sigma\sigma'} \Bigr) \psi_{\sigma' n'} (i\omega) \nn &&
- \beta \int d^{d} \bm{r} \frac{3}{2 U} [\boldsymbol{\Phi}(\boldsymbol{r})]^{2} \Bigr] ,
\eqa
where the magnetization order parameter is assumed to be static and determined self-consistently in the mean-field analysis. The Gaussian integral for $\psi_{\sigma n}(i\omega)$ gives rise to the following expression for the mean field free energy
\bqa
&& \mathcal{F} \approx - T \int_{-\infty}^{\infty} d v(\bm{r}) P[v(\bm{r})] \nn && \sum_{n} \Bigl[
\ln \Bigl\{ 1 + \exp\Bigl( - \frac{\varepsilon_{n} - \int d^{d} \bm{r} |\Psi_{n}(\bm{r})|^{2} {\Phi}(\bm{r})}{T} \Bigr) \Bigr\} \nn
&& + \ln \Bigl\{ 1 + \exp\Bigl( - \frac{\varepsilon_{n} + \int d^{d} \bm{r} |\Psi_{n}(\bm{r})|^{2} {\Phi}(\bm{r})}{T} \Bigr) \Bigr\} \nn && - \frac{1}{T} \int d^{d} \bm{r} \frac{3}{2 U} [{\Phi}(\boldsymbol{r})]^{2} \Bigr] , \label{Effective_Free_Energy} \eqa where the magnetization order parameter $\bm{\Phi}(\bm{r}) = \Phi(\bm{r}) \bm{\hat{z}}$ is determined by the self-consistent equation \bqa && \frac{3}{U} {\Phi}(\bm{r}) = \sum_{n} |\Psi_{n}(\bm{r})|^{2} \Bigl\{ f\Bigl(\varepsilon_{n} - \int d^{d} \bm{r} |\Psi_{n}(\bm{r})|^{2} {\Phi}(\bm{r})\Bigr) \nn && - f\Bigl(\varepsilon_{n} + \int d^{d} \bm{r} |\Psi_{n}(\bm{r})|^{2} {\Phi}(\bm{r})\Bigr) \Bigr\} . \label{Self_Consistent_Equation} \eqa $f(\varepsilon_{n}) = \frac{1}{e^{\varepsilon_{n}/T} + 1}$ is the Fermi-Dirac distribution function.

It should be noted that coupling effects between $n$ and $n'$ are neglected as the zeroth order approximation. We point out that the existence of off-diagonal terms in the energy space is a general feature in any mean-field theories with strong randomness. Indeed, this approximation has been used not only in the Stoner-Anderson problem but also in the Kondo-Anderson transition \cite{PTK2}. In order to justify the diagonal-in-energy approximation, one can consider higher order processes such as $\langle e^{- \mathcal{S}_{int} } \rangle_{0} \approx \exp\Big\{ - \langle \mathcal{S}_{int} \rangle_{0} + \frac{1}{2} \Big( \langle \mathcal{S}_{int}^{2} \rangle_{0} - \langle \mathcal{S}_{int} \rangle_{0}^{2} \Big) \Big\}$ with $\mathcal{S}_{int} = \sum_{i \omega} \sum_{n \not= n'} \int d^{d} \bm{r} \Psi_{n}^{\dagger}(\bm{r}) \Psi_{n'}(\bm{r}) \Phi(\bm{r}) \sigma \psi_{\sigma n}^{\dagger}(i\omega) \psi_{\sigma n'}(i\omega)$, where \bqa && \langle \mathcal{O} \rangle_{0} = \frac{1}{Z}_{0} \int D \psi_{\sigma n}(i\omega) \mathcal{O} e^{- \mathcal{S}_{0}} , \nn && \mathcal{S}_{0} = \sum_{i \omega} \sum_{n} \psi_{\sigma n}^{\dagger}(i \omega) \Big(- i \omega + \varepsilon_{n} \nn && + \int d^{d} \bm{r} |\Psi_{n}(\bm{r})|^{2} \sigma \Phi(\bm{r}) \Big) \psi_{\sigma n}(i\omega) , \nn && Z_{0} = \int D \psi_{\sigma n}(i\omega) e^{- \mathcal{S}_{0}} , \nonumber \eqa respectively. While the first-order term of $\langle \mathcal{S}_{int} \rangle_{0}$ vanishes identically, the second-order term can be expressed as follows \bqa && \mathcal{S}^{(2)}_{eff} \equiv - \frac{1}{2} \Big( \langle \mathcal{S}_{int}^{2} \rangle_{0} - \langle \mathcal{S}_{int} \rangle_{0}^{2} \Big) \nn && = - \frac{1}{2} \sum_{i \omega} \sum_{i \omega'} \sum_{n \not= n'} \sum_{m \not= m'} \int d^{d} \bm{r} \int d^{d} \bm{r}' \nn && \Psi_{n}^{\dagger}(\bm{r}) \Psi_{n'}(\bm{r}) \Psi_{m}^{\dagger}(\bm{r}') \Psi_{m'}(\bm{r}') \Phi(\bm{r}) \sigma \Phi(\bm{r}') \sigma' \nn && \psi_{\sigma n}^{\dagger}(i\omega) \langle \psi_{\sigma n'}(i\omega) \psi_{\sigma' m}^{\dagger}(i\omega') \rangle_{c} \psi_{\sigma' m'}(i\omega') \nn && = - \frac{1}{2} \sum_{i \omega} \sum_{n \not= n'} \sum_{m'} \int d^{d} \bm{r} \int d^{d} \bm{r}' \nn && \Psi_{n}^{\dagger}(\bm{r}) \Psi_{n'}(\bm{r}) \Psi_{n'}^{\dagger}(\bm{r}') \Psi_{m'}(\bm{r}') \Phi(\bm{r}) \Phi(\bm{r}') \nn && \psi_{\sigma n}^{\dagger}(i\omega) \frac{1}{- i \omega + \varepsilon_{n'} + \int d^{d} \bm{r} |\Psi_{n'}(\bm{r})|^{2} \sigma \Phi(\bm{r}) } \psi_{\sigma m'}(i\omega) \nn && \approx - \frac{1}{2} \sum_{i \omega} \sum_{n} \psi_{\sigma n}^{\dagger}(i\omega) \psi_{\sigma n}(i\omega) \nn && \sum_{n'}\frac{\int d^{d} \bm{r} \int d^{d} \bm{r}' \Psi_{n}^{\dagger}(\bm{r}) \Psi_{n'}(\bm{r}) \Psi_{n'}^{\dagger}(\bm{r}') \Psi_{n}(\bm{r}') \Phi(\bm{r}) \Phi(\bm{r}')}{- i \omega + \varepsilon_{n'} + \int d^{d} \bm{r} |\Psi_{n'}(\bm{r})|^{2} \sigma \Phi(\bm{r}) } . \nonumber \eqa Here, $\langle \cdot\cdot\cdot \rangle_{c}$ means to keep the connected part of diagrams. We point out that the diagonal approximation has been used again in the last step. Then, this diagonal approximation can be justified when the following condition is satisfied \bqa && \frac{\frac{1}{2} \Big| \sum_{n'} \frac{\int d^{d} \bm{r} \int d^{d} \bm{r}' \Psi_{n}^{\dagger}(\bm{r}) \Psi_{n'}(\bm{r}) \Psi_{n'}^{\dagger}(\bm{r}') \Psi_{n}(\bm{r}') \Phi(\bm{r}) \Phi(\bm{r}')}{- i \omega + \varepsilon_{n'} + \int d^{d} \bm{r} |\Psi_{n'}(\bm{r})|^{2} \sigma \Phi(\bm{r}) } \Big|}{\Big| \int d^{d} \bm{r} |\Psi_{n}(\bm{r})|^{2} \sigma \Phi(\bm{r}) \Big|} \ll 1 . \nonumber \eqa We claim that this criterion is fulfilled when we evaluate the critical temperature, where the local magnetization order parameter of a rare region vanishes. This explains why the diagonal approximation also works in the Kondo-Anderson problem for the distribution function of the Kondo temperature.

\subsubsection{Eigenfunction multifractality}

Next, we replace the integral for the average in disorder configurations $\int_{-\infty}^{\infty} d v(\bm{r}) P[v(\bm{r})]$ with $\int_{-\infty}^{\infty} \Pi_{n} d \alpha_{n}(\boldsymbol{r}) P[\{\alpha_{n}(\boldsymbol{r})\}]$ for the average in the statistics of eigenfunctions. We note that all the information for the statistics of eigenfunctions are encoded into the distribution function of $P[\{\alpha_{n}(\boldsymbol{r})\}]$ with $\alpha_{n}(\bm{r}) = - \frac{\ln |\Psi_{n}(\bm{r})|^{2}}{\ln L}$ \cite{Alpha}, given by the Gaussian distribution function for all $\alpha_{n}(\boldsymbol{r})$ \cite{Multifractality}, which will be clarified below. An important point is how to perform the integration for the wave-function distribution. Recently, this procedure has been discussed intensively, where an idea is to take into account the so-called joint distribution function which deals with pairs of eigenfunctions \cite{PTK2}, given by
\bqa
&& \int_{-\infty}^{\infty} \Pi_{n} d \alpha_{n}(\boldsymbol{r}) P[\{\alpha_{n}(\boldsymbol{r})\}]
\approx \int_{-\infty}^{\infty} d \alpha (\boldsymbol{r}) P^{(1)}[\alpha (\boldsymbol{r})] \nn &&
\int_{-\infty}^{\infty} \Pi_{n} d \alpha_{n}(\boldsymbol{r}) \frac{P^{(2)}[\alpha_{n}(\boldsymbol{r})
\not= \alpha(\boldsymbol{r})]}{P^{(1)}[\alpha(\boldsymbol{r})]} .
\eqa
Here, $P^{(2)}[\alpha_{n}(\boldsymbol{r}) \not= \alpha(\boldsymbol{r})]$ is the joint distribution function, where one eigenfunction $\alpha(\boldsymbol{r})$ is at the mobility edge $\varepsilon_{m}$ and the other wave function $\alpha_{n}(\boldsymbol{r})$ is away from the mobility edge. On the other hand, $P^{(1)}[\alpha (\boldsymbol{r})]$ is the distribution function for the single eigenfunction at the mobility edge. Both distribution functions are given by the Gaussian distribution function (the log-normal distribution function for the intensity of an eigenfunction), constructed to reproduce the wave-function multifractality of the random matrix theory or the supersymmetric nonlinear $\sigma-$model approach \cite{Multifractality}. Then, this integral means to perform the integral for $\alpha_{n}(\boldsymbol{r})$ with a fixed $\alpha(\boldsymbol{r})$ first, based on the mutual distribution function, and to do for $\alpha(\boldsymbol{r})$ next, based on the single eigenfunction distribution function. As a result, we reach the following expression
\begin{widetext}
\bqa
&& \mathcal{F} \approx - T \int_{-\infty}^{\infty} d \alpha (\boldsymbol{r}) P^{(1)}[\alpha (\boldsymbol{r})]
\sum_{n} \Bigl[ \ln \Bigl\{ 1 + \exp\Bigl( - \frac{\varepsilon_{n} - \int d^{d} \bm{r} \Bigl\langle |\Psi_{n}(\bm{r})|^{2}
\Bigr\rangle_{|\Psi_{m}(\bm{r})|^{2} = L^{-\alpha(\boldsymbol{r})}} {\Phi}(\bm{r})}{T} \Bigr) \Bigr\} \nn
&& + \ln \Bigl\{ 1 + \exp\Bigl( - \frac{\varepsilon_{n} + \int d^{d} \bm{r} \Bigl\langle |\Psi_{n}(\bm{r})|^{2}
\Bigr\rangle_{|\Psi_{m}(\bm{r})|^{2} = L^{-\alpha(\boldsymbol{r})}} {\Phi}(\bm{r})}{T} \Bigr) \Bigr\} \Bigr]
+ \int_{-\infty}^{\infty} d \alpha(\boldsymbol{r}) P^{(1)}[\alpha(\boldsymbol{r})] \int d^{d} \bm{r} \frac{3}{2 U} [{\Phi}(\boldsymbol{r})]^{2} ,
\label{Free_Energy_Mother}
\eqa
\end{widetext}
where the average for the mutual distribution function gives rise to
\bqa
&& \Bigl\langle |\Psi_{n}(\bm{r})|^{2} \Bigr\rangle_{|\Psi_{m}(\bm{r})|^{2} = L^{-\alpha(\boldsymbol{r})}}
\nn && \equiv \int_{-\infty}^{\infty} \Pi_{n} d \alpha_{n}(\boldsymbol{r}) \frac{P^{(2)}[\alpha_{n}(\boldsymbol{r})
\not= \alpha(\boldsymbol{r})]}{P^{(1)}[\alpha(\boldsymbol{r})]} |\Psi_{n}(\bm{r})|^{2} \nn &&
= L^{-d} \Bigl| \frac{\varepsilon_{n} - \varepsilon_{m}}{\varepsilon_{c}} \Bigr|^{r_{\alpha(\boldsymbol{r})}} \label{Average_Multifractality}
\eqa
with an exponent \cite{PTK2}
\bqa
&& r_{\alpha(\boldsymbol{r})} = \frac{\alpha(\boldsymbol{r}) - \alpha_{0}}{d} - \frac{\eta}{2d} g_{nm} ,
~~~ \eta = 2 (\alpha_{0} - d) \\ &&
g_{nm} = \frac{\ln |(\varepsilon_{n} - \varepsilon_{m})/\varepsilon_{c}|}{d \ln L}  .
\eqa
Here, $L$ is the size of a system, $d$ is a space dimension, $\varepsilon_{c}$ is a cutoff, which shows strong correlations of eigenfunctions with different energies up to $\varepsilon_{c}$, and $\alpha_{0}$ is a typical value of the logarithm of an eigenfunction.

Taking the integral for discrete energies as follows $\sum_{n} \approx \rho_{m} \int_{-\varepsilon_{c}}^{\varepsilon_{c}} d \varepsilon_{n}$ \cite{Energy_Level_Comment}, we obtain
\bqa
&& \mathcal{F} \approx - T \rho_{m} \int_{-\varepsilon_{c}}^{\varepsilon_{c}} d \varepsilon_{n}
\int_{-\infty}^{\infty} d \alpha (\boldsymbol{r}) P^{(1)}[\alpha (\boldsymbol{r})] \nn &&
\Bigl[ \ln \Bigl\{ 1 + \exp\Bigl( - \frac{\varepsilon_{n} - \Delta_{n}}{T} \Bigr) \Bigr\}
+ \ln \Bigl\{ 1 + \exp\Bigl( - \frac{\varepsilon_{n} + \Delta_{n}}{T} \Bigr) \Bigr\} \Bigr] \nn
&& + \int_{-\infty}^{\infty} d \alpha(\boldsymbol{r}) P^{(1)}[\alpha(\boldsymbol{r})]
\int d^{d} \bm{r} \frac{3}{2 U} [{\Phi}(\boldsymbol{r})]^{2}
\label{Free_Energy_Inhomogeneity}
\eqa
with $\Delta_{n} \equiv \int d^{d} \bm{r} \Bigl\langle |\Psi_{n}(\bm{r})|^{2} \Bigr\rangle_{|\Psi_{m}(\bm{r})|^{2} = L^{-\alpha(\boldsymbol{r})}} {\Phi}(\bm{r}) \approx L^{-d} \int d^{d} \bm{r} \Bigl| \frac{\varepsilon_{n}}{\varepsilon_{c}} \Bigr|^{r_{\alpha(\boldsymbol{r})}} \Phi(\bm{r})$.
The magnetization order parameter is determined by the self-consistent equation for a given function $\alpha(\bm{r})$
\bqa
&& \Delta_{l} = \frac{U}{3} \rho_{m} \int \frac{d^{d} \bm{r} }{L^d} \Bigl| \frac{\varepsilon_{l}}{\varepsilon_{c}}
\Bigr|^{r_{\alpha(\bm{r})}} \nn && \int_{-\varepsilon_{c}}^{\varepsilon_{c}} d \varepsilon_{n}
\Bigl| \frac{\varepsilon_{n}}{\varepsilon_{c}} \Bigr|^{r_{\alpha(\boldsymbol{r})}} \Bigl\{ f(\varepsilon_{n} - \Delta_n)
- f(\varepsilon_{n} + \Delta_n) \Bigr\} .
\eqa
This self-consistent equation shows correlation effects in the energy space. In order to obtain $\Delta_{l}$, we should know $\Delta_{n}$ for all values of $n$, exhibiting correlations in the energy space. Solving these coupled equations in the energy space, we obtain the magnetization order parameter in a given function of $\alpha(\bm{r})$. Performing the average for $\alpha(\bm{r})$ with an appropriate distribution function $P[\alpha(\bm{r})]$, we take into account correlation effects in the energy space.

\subsubsection{Local mean-field theory}

Unfortunately, this effective free energy is not completely local in space since there exists an integral for the whole space given by $\int d^{d} \bm{r} \Bigl\langle |\Psi_{n}(\bm{r})|^{2} \Bigr\rangle_{|\Psi_{m}(\bm{r})|^{2} = L^{-\alpha(\boldsymbol{r})}} {\Phi}(\bm{r})$. An essential simplification is to ``lose" or ``overestimate" (more precisely, see the below discussion) the information on strong spatial inhomogeneity as the zeroth order approximation. Replacing $\alpha(\bm{r})$ with $\alpha$ and taking the integral for discrete energies as follows $\sum_{n} \approx \rho_{m} \int_{-\varepsilon_{c}}^{\varepsilon_{c}} d \varepsilon_{n}$ in Eq. (\ref{Free_Energy_Mother}), we obtain a local mean-field theory for the Stoner transition at Anderson localization
\bqa
&& L^{-d} \mathcal{F} \approx - T \rho_{m} \int_{-\varepsilon_{c}}^{\varepsilon_{c}} d \varepsilon_{n}
\int_{-\infty}^{\infty} d \alpha P(\alpha) \nn && \Bigl[ \ln \Bigl\{ 1 + \exp\Bigl( - \frac{\varepsilon_{n}
- \Bigl| \frac{\varepsilon_{n}}{\varepsilon_{c}} \Bigr|^{r_{\alpha}} \Phi(\alpha)}{T} \Bigr) \Bigr\} \nn &&
+ \ln \Bigl\{ 1 + \exp\Bigl( - \frac{\varepsilon_{n}
+ \Bigl| \frac{\varepsilon_{n}}{\varepsilon_{c}} \Bigr|^{r_{\alpha}} \Phi(\alpha)}{T} \Bigr) \Bigr\} \Bigr] \nn
&& + \int_{-\infty}^{\infty} d \alpha P(\alpha) \frac{3}{2 U} \Phi^{2}(\alpha) ,
\eqa
where $L^{-d} \int d^{d} \bm{r} {\Phi}(\bm{r})$ is replaced with $\Phi(\alpha)$, determined by the ``gap" equation for the order parameter
\bqa
&& \Phi(\alpha) = \frac{U}{3} \rho_{m} \int_{-\varepsilon_{c}}^{\varepsilon_{c}} d \varepsilon_{n}
\Bigl| \frac{\varepsilon_{n}}{\varepsilon_{c}} \Bigr|^{r_{\alpha}} \Bigl\{ f\Bigl(\varepsilon_{n}
- \Bigl| \frac{\varepsilon_{n}}{\varepsilon_{c}} \Bigr|^{r_{\alpha}} \Phi(\alpha) \Bigr) \nn &&
- f\Bigl(\varepsilon_{n} + \Bigl| \frac{\varepsilon_{n}}{\varepsilon_{c}} \Bigr|^{r_{\alpha}} \Phi(\alpha) \Bigr) \Bigr\} .
\label{Gap_Equation_FM}
\eqa
The distribution function is given by
\bqa
&& P(\alpha) = \mathcal{N} L^{ - \frac{(\alpha-\alpha_{0})^{2}}{2\eta}} ,
\eqa
where $\mathcal{N}$ is a positive numerical constant determined from $\int_{-\infty}^{\infty} d \alpha P(\alpha) = 1$.

The critical temperature for a given disorder configuration is determined by \bqa && 1 = \frac{U \rho_{m}}{3 T_{c}} \int_{0}^{\varepsilon_{c}} d \varepsilon \Bigl( \frac{\varepsilon}{\varepsilon_{c}} \Bigr)^{2 r_{\alpha}} \frac{1}{\cosh^{2}\Bigl(\frac{\varepsilon}{2T_{c}}\Bigr)} , \label{Local_Tc} \eqa which results from Eq. (\ref{Gap_Equation_FM}), performing the Taylor expansion for the order parameter up to the first order in the right hand side. Then, we obtain $T_{c} = T_{c}(r_{\alpha})$.
%
%
This relation allows us to translate $P(\alpha)$ into $P(T_c) = \left|\frac{dT_c}{d \alpha} \right|^{-1} P(\alpha)$.

We would like to point out that the magnetization order parameter is given by a function of $\alpha$ and both the $\alpha-$dependent $\Phi(\alpha)$ and the integration for $\alpha$ with $P(\alpha)$ are expected to keep correlation
effects in the energy space. However, it is true that strong spatial fluctuations in the intensity of eigenfunctions are overestimated in our mean-field theory. A physical picture for this mean-field analysis is as follows. Suppose an island at position $\bm{r}$ with a characteristic length scale, determined by both interactions and disorders, where the intensity of an eigenfunction may be regarded to be uniform, responsible for the uniform magnetization within the island. Then, we consider another island at position $\bm{r}'$ near the previous island, introducing some couplings such as electron hopping and magnetic interaction between these nearest-neighbor islands. Based on this granular picture, one may perform a weak-coupling analysis for interactions between these granules. One may suspect three kinds of possibilities, which correspond to relevance, irrelevance, and marginality of granular interactions, respectively. We believe that the present mean-field analysis focuses on the case of irrelevant granular interactions, giving rise to random magnetization for each intensity of eigenfunctions beyond a certain (granular) length scale \cite{Discussion_Inhomogeneity}.
%
%

The above discussion can be stated more mathematically as follows. Suppose two competing length scales: One is the length scale referred to as the size of a granule, allowing us to replace $\alpha(\bm{r})$ with $\alpha$, and the other is the correlation length of the magnetization order parameter to guarantee uniformity within the length scale. The local mean-field theory can be justified when the first length scale is larger than the second. In order to verify whether this is possible or not, let us consider the other case that the second length scale is larger than the first. Then, we are allowed to set $\Delta_{n} \approx \Big( \int \frac{d^{d} \bm{r} }{L^d} \Bigl| \frac{\varepsilon_{n}}{\varepsilon_{c}} \Bigr|^{r_{\alpha(\boldsymbol{r})}} \Big) \Phi$. As a result, the self-consistent equation for the order parameter is given by
\bqa
&& \Big( \int \frac{d^{d} \bm{r} }{L^d} \Bigl| \frac{\varepsilon_{l}}{\varepsilon_{c}} \Bigr|^{r_{\alpha(\boldsymbol{r})}} \Big) \Phi = \frac{U}{3} \rho_{m} \int \frac{d^{d} \bm{r} }{L^d} \Bigl| \frac{\varepsilon_{l}}{\varepsilon_{c}} \Bigr|^{r_{\alpha(\bm{r})}} \nn && \int_{-\varepsilon_{c}}^{\varepsilon_{c}} d \varepsilon_{n}
\Bigl| \frac{\varepsilon_{n}}{\varepsilon_{c}} \Bigr|^{r_{\alpha(\boldsymbol{r})}} \Bigl\{ f\Big(\varepsilon_{n} - \int \frac{d^{d} \bm{r} }{L^d} \Bigl| \frac{\varepsilon_{n}}{\varepsilon_{c}} \Bigr|^{r_{\alpha(\boldsymbol{r})}} \Phi\Big) \nn && - f\Big(\varepsilon_{n} + \int \frac{d^{d} \bm{r} }{L^d} \Bigl| \frac{\varepsilon_{n}}{\varepsilon_{c}} \Bigr|^{r_{\alpha(\boldsymbol{r})}} \Phi\Big) \Bigr\} . \nonumber
\eqa
It is not easy to see the existence of a solution. In this respect, performing the Taylor expansion for the order parameter up to the first order in the right hand side, we obtain an equation for the critical temperature, given by \bqa && \int \frac{d^{d} \bm{r} }{L^d} \Bigl| \frac{\varepsilon_{l}}{\varepsilon_{c}} \Bigr|^{r_{\alpha(\boldsymbol{r})}} \nn && = \frac{U \rho_{m}}{3 T_{c}} \int \frac{d^{d} \bm{r} }{L^d} \Bigl| \frac{\varepsilon_{l}}{\varepsilon_{c}} \Bigr|^{r_{\alpha(\bm{r})}} \int_{0}^{\varepsilon_{c}} d \varepsilon \Bigl( \frac{\varepsilon}{\varepsilon_{c}} \Bigr)^{2 r_{\alpha(\boldsymbol{r})}} \frac{1}{\cosh^{2}\Bigl(\frac{\varepsilon}{2T_{c}}\Bigr)} . \nonumber \eqa Given $\alpha(\bm{r})$, is there a solution of this equation? Although we do not know the answer in a general situation, we know that there is a solution when $\alpha(\bm{r})$ is replaced with $\alpha$ in this equation. We stress that both length scales should be determined self-consistently beyond the present theoretical consideration. This granular-medium picture deserves to be investigated more sincerely near future.

\subsection{The Kondo effect}

\subsubsection{Formulation}

We start from an effective Kondo Hamiltonian
\bqa
&& \mathcal{H} = \int d^{d} \boldsymbol{r} \Big\{  c^\dagger_\sigma (\tau, \bm r) \left(- \frac{\boldsymbol{\nabla_{\boldsymbol{r}}}^{2}}{2m}
- \mu + v(\bm r)\right) c_\sigma (\tau, \bm r) \nn && + J_K \delta^{(d)}(\bm{r}) \bm s \cdot \bm S \Big\}
\eqa
where the spin of the conduction electron is given as $\bm s = c^\dagger_\sigma (\tau, \bm r) \bm \sigma_{\sigma \sigma'} c_{\sigma'} (\tau, \bm r)$ and the spin of the impurity as $\bm S = f^\dagger_\sigma (\tau) \bm \sigma_{\sigma \sigma'} f_{\sigma'} (\tau)$ in the fermion projective representation, backup by the single occupancy constraint $f_{\sigma}^{\dagger} f_{\sigma} = N_{s} S$ with $N_{s} = 2$ and $S = 1/2$ \cite{Hewson_Kondo}.

Performing the Hubbard-Stratonovich transformation for the Kondo-hybridization spin-singlet channel, we obtain the following expression for the free energy
\bqa
&& \mathcal{F} = -\frac{1}{\beta} \int_{-\infty}^{\infty} dv(\bm r) P[v(\bm{r})] \ln \int
D c_\sigma(\tau, \bm r) D f_\sigma (\tau) \nn && D b(\tau) D \lambda(\tau)
\exp \Bigl[-\int_0^\beta d\tau \Bigl\{ \int d^d \bm r c^\dagger_\sigma (\tau, \bm r) \Bigl(\partial_\tau - \mu
\nn && - \frac{\boldsymbol{\nabla_{\boldsymbol{r}}}^{2}}{2m} + v(\bm r) \Bigr) c_\sigma (\tau, \bm r)
- \frac{J_K}{N_s} \Bigl(c^\dagger_\sigma (\tau) b^\dagger_\sigma (\tau) f_\sigma (\tau) \nn && + H.c. \Bigr)
+ \frac{J_K}{N_s} b^\dagger_\sigma (\tau) b_\sigma (\tau) +
f^\dagger_\sigma (\tau) \partial_\tau f_\sigma (\tau) \nn &&
+ i \lambda (\tau) \Bigl( f^\dagger_\sigma (\tau) f_\sigma (\tau) - N_s S \Bigr) \Bigr\} \Bigr] ,
\eqa
where the disorder average is taken into account. $b_\sigma (\tau)$ may be identified with an order parameter for the local Fermi-liquid state, given by $b_\sigma (\tau) = \Bigl\langle c^\dagger_\sigma (\tau) f_\sigma (\tau) \Bigr\rangle$ in the saddle-point approximation \cite{Hewson_Kondo}. $\lambda(\tau)$ is a Lagrange multiplier field to impose the fermion-number constraint.

\subsubsection{Eigenfunction multifractality}

Following the same procedure as that of the previous subsection, we rewrite this effective free energy in terms of the eigenfunction for a given disorder configuration [Eqs. (\ref{Eigenfunction}) and (\ref{Fourier_Disorder})]. As a result, we obtain
\bqa
&& \mathcal{F} = - \frac{1}{\beta} \int_{-\infty}^{\infty} \Pi_{n} d \alpha_{n}(\boldsymbol{r}) P[\{\alpha_{n}(\boldsymbol{r})\}]
\ln \int D \psi_{\sigma n}(\tau) D f_\sigma (\tau) \nn && D b(\tau) D \lambda(\tau) \exp \Bigl[-\int_0^\beta d\tau \Bigl\{ \sum_n \psi^\dagger_{\sigma n}(\tau) \Bigl(\partial_\tau + \varepsilon_n \Bigr) \psi_{\sigma n}(\tau)
\nn && - \frac{J_K}{N_s} \Bigl(\sum_n \Psi_n^{\dagger} \psi^\dagger_{\sigma n}(\tau) b^\dagger_\sigma (\tau) f_\sigma (\tau) + H.c. \Bigr)
+ \frac{J_K}{N_s} b^\dagger_\sigma (\tau) b_\sigma (\tau) \nn && + f^\dagger_\sigma (\tau) \partial_\tau f_\sigma (\tau)
+ i \lambda (\tau) \Bigl( f^\dagger_\sigma (\tau) f_\sigma (\tau) - N_s S \Bigr) \Bigr\} \Bigr].
\eqa

Taking into account the mean-field approximation of $b(\tau) \to b$ and $\lambda(\tau) \to - i \lambda$ and performing both Gaussian integrals for conduction electrons and localized fermions, we obtain
%
%
\bqa
&& \mathcal{F} = -\frac{1}{\beta} \int_{-\infty}^{\infty} \Pi_{n} d \alpha_{n}(\boldsymbol{r}) P[\{\alpha_{n}(\boldsymbol{r})\}] \nn &&
\Bigl\{N_s \sum_{i\omega} \ln \Bigl( -i \omega + \lambda  + \frac{J_K^2 b^2}{N_s^2} \sum_n \frac{|\Psi_n|^2}{i \omega - \varepsilon_n} \Bigr) \nn &&
-\beta \Bigl(\frac{J_K}{N_s} b^2 - N_s S \lambda \Bigr) + N_s \sum_n \ln(1 + e^{-\beta \varepsilon_n}) \Bigr\} .
\eqa
Minimizing the free energy with respect to $\lambda$ and $b$, respectively, yields
\bqa
&& S = \frac{1}{\beta} \sum_{i\omega} \frac{1}{i \omega - \lambda - \frac{J_K^2 b^2}{N_s^2} \sum_n \frac{|\Psi_n|^2}{i \omega - \varepsilon_n}} ,
\label{dF_dlambda}\\
&& 1 = -\frac{J_K}{\beta} \sum_{i \omega} \frac{\sum_n \frac{|\Psi_n|^2}{i \omega - \varepsilon_n}}
{i \omega - \frac{J_K^2 b^2}{N_s^2} \sum_n \frac{|\Psi_n|^2}{i \omega - \varepsilon_n}} .
\label{dF_db}
\eqa

It is straightforward to determine the chemical potential $\lambda$ of localized fermions. Substituting
\bqa
&& \sum_n \frac{|\Psi_n|^2}{i \omega - \varepsilon_n} = -i\omega \int_{-\infty}^{\infty} d\varepsilon_n \rho(\varepsilon_n)
\frac{|\Psi_n|^2}{\omega^2 + \varepsilon_n^2} \nn && \approx -i\pi \rho_\omega |\Psi_\omega|^2 \mbox{sgn}(\omega)
\eqa
to Eq. (\ref{dF_dlambda}) gives $\lambda = 0$, where the last approximation takes the low-frequency limit. Zero chemical potential means that localized fermions are at half filling, i.e., in the Kondo regime.

\subsubsection{Local mean-field theory}

Compared with the Stoner transition, the local mean-field theory is quite natural in the Kondo effect since the ``phase transition" itself is local. The position may be regarded as a dummy variable. Taking into account $|\Psi_n|^2 \longrightarrow \langle |\Psi_n|^2\rangle$ with $\alpha(\bm r) \to \alpha$ and $\sum_{n} \to \rho_{m} \int_{-\varepsilon_{c}}^{\varepsilon_{c}} d \varepsilon_{n}$, we arrive at a local mean-field theory for the Kondo effect at the Anderson transition
\bqa
&& \mathcal{F} = -\frac{\rho_m}{\beta} \int_{-\varepsilon_c}^{\varepsilon_c}  d \varepsilon_{n} \int_{-\infty}^{\infty} d \alpha P(\alpha) \nn &&
\Bigl\{N_s \sum_{i\omega} \ln \Bigl( -i \omega + \frac{J_K^2 b^2}{N_s^2} \sum_n \frac{\left|\frac{\varepsilon_n}{\varepsilon_c}\right|^{r_\alpha}}{i \omega - \varepsilon_n} \Bigr)
- \beta \frac{J_K}{N_s} b^2 \nn && + N_s \sum_n \ln(1 + e^{-\beta \varepsilon_n}) \Bigr\} .
\eqa
As a result, the Kondo temperature is determined by
\bqa
&& 1 = \frac{J_K \rho_m}{2} \int_{-\varepsilon_c}^{\varepsilon_c}  d \varepsilon_{n} \left|\frac{\varepsilon_n}{\varepsilon_c}\right|^{r_\alpha} \frac{1}{\varepsilon_n} \tanh \left(\frac{\varepsilon_n}{2 T_{K}} \right) ,
\eqa
essentially the same as that of Refs. \cite{PTK1,PTK2}. Approximating $\tanh x \approx x$ for $x < 1$ and $\tanh x \approx 1$ for $x >1$, we find
\bqa
T_K & = & \frac{\varepsilon_c}{2}\left[(r_\alpha + 1) \left(1 -\frac{r_\alpha}{J_{K} \rho} \right) \right]^{1/r_\alpha}
\eqa
for $T_K < \frac{\varepsilon_c}{2}$ and $-1 < r_\alpha < J_{K} \rho$, and
\bqa
T_K & = & \frac{\varepsilon_c}{2} \frac{J_{K} \rho}{r_\alpha + 1}
\eqa
for $T_K > \frac{\varepsilon_c}{2}$ and $r_\alpha > -1$, respectively.
%
%
This relation determines the distribution of the Kondo temperature via $P(T_K) = \left|\frac{dT_K}{d \alpha} \right|^{-1} P(\alpha)$.

We would like to point out that our way how to obtain the distribution function of the Kondo temperature is not the same as that in Ref. \cite{PTK2} although the same mean-field equation is utilized. When we find the distribution function of the Kondo temperature from that of disorder-eigenfunctions, an essential point is how to introduce the constraint of the mean-field equation for the critical temperature into the equation of the distribution function. More rigorously speaking, the problem is how to perform the integration of the Lagrange multiplier field $t$ for the delta-function imposing the constraint of the mean-field equation, given by \bqa && P(T_{K}) = - \int_{0}^{\infty} d \alpha P(\alpha) \frac{d F[T_{K}]}{d T_{K}} \delta(1 - F[T_{K}]) \nn && = - \int_{0}^{\infty} d \alpha P(\alpha) \frac{d }{d T_{K}} \int_{-\infty}^{\infty} d t \frac{i e^{i t}}{2 \pi t} \exp\big(- i t F[T_{K}] \big) , \nonumber \eqa where $F[T_{K}] = \frac{J_K \rho_m}{2} \int_{-\varepsilon_c}^{\varepsilon_c}  d \varepsilon_{n} \left|\frac{\varepsilon_n}{\varepsilon_c}\right|^{r_\alpha} \frac{1}{\varepsilon_n} \tanh \left(\frac{\varepsilon_n}{2 T_{K}} \right)$ is the right hand side of the mean-field equation for the Kondo temperature \cite{PTK2}. The previous study performs the integration for $t$ up to the second order analytically. On the other hand, we perform the integration for $t$ up to an infinite order numerically.

\section{A criterion of quantum Griffiths phenomena vs. smeared phase transitions}

\begin{figure}[t]
\includegraphics[width=0.45\textwidth]{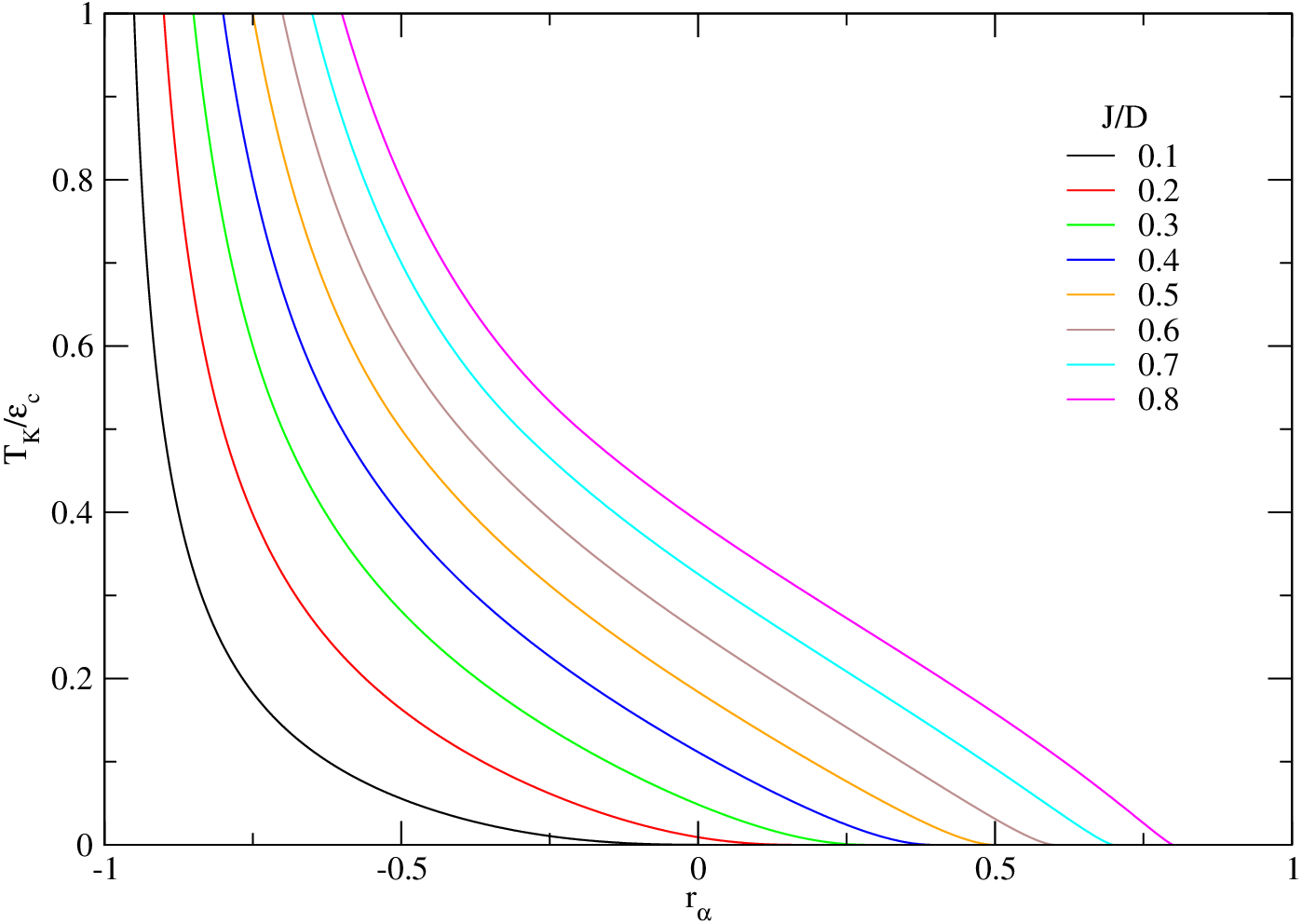}
\caption{Local Kondo temperature vs. $r_\alpha$ for various Kondo interactions. The local Kondo temperature as a function of $r_{\alpha}$ can be found from Eq. (27). A noticeable point is that the local Kondo temperature is well defined to decrease continuously in the pseudogap region until it vanishes. It turns out that the hybridization order parameter decreases to vanish continuously, exhibiting a conventional ``second-order transition" in the phase diagram of the Kondo coupling and temperature with a given $r_{\alpha}$, not shown here.}
\label{Kondo_TKr}
\vspace*{1cm}
\includegraphics[width=0.45\textwidth]{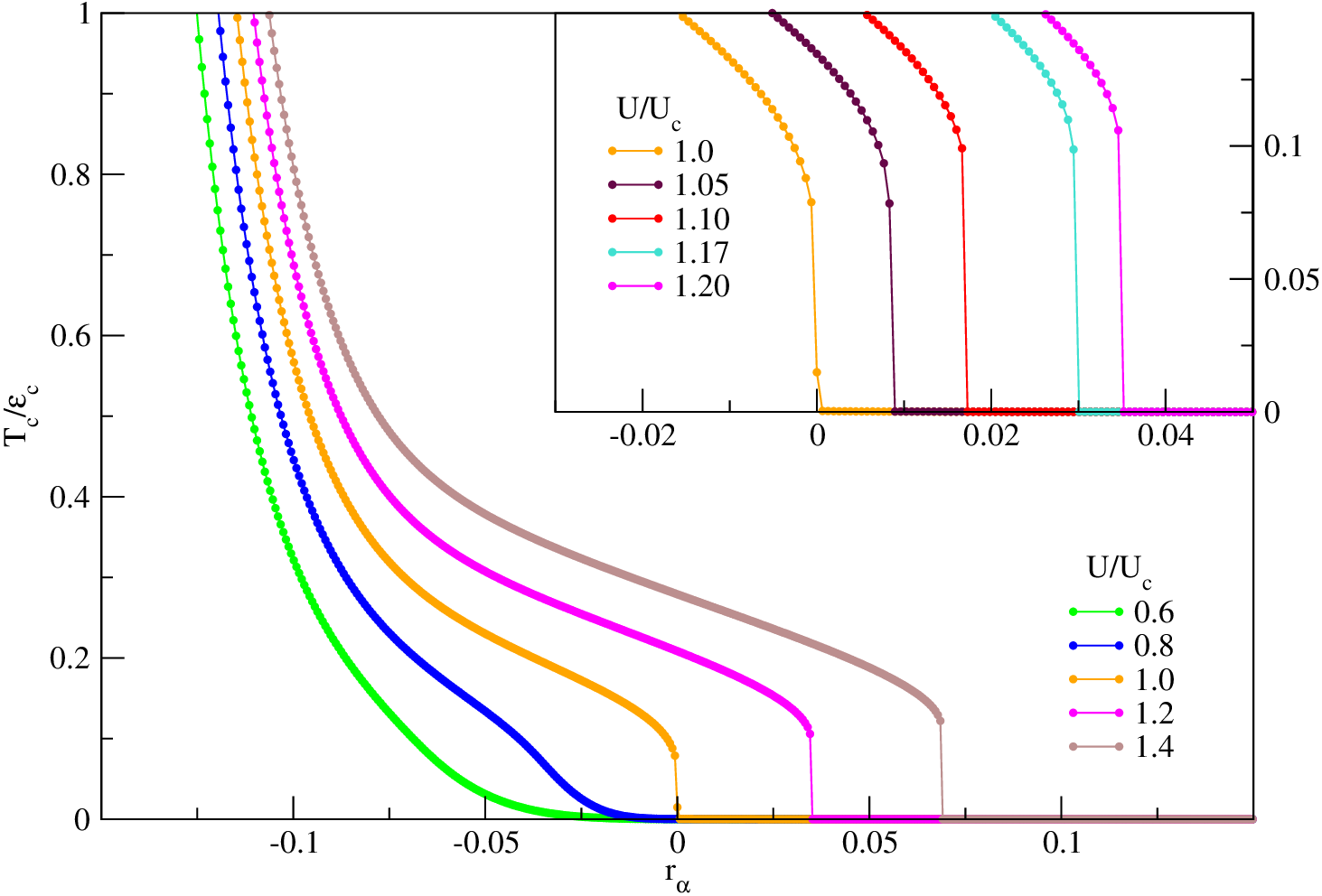}
\caption{Local critical temperature of the Stoner transition vs. $r_\alpha$ for various local interactions. The local critical temperature as a function of $r_{\alpha}$ can be found from Eq. (18). The local critical temperature drops down abruptly above a certain positive value of $r_{\alpha}$, i.e., in the pseudogap region with a fixed interaction parameter, where the inset confirms this observation. This implies that the local pseudogap region shows an abrupt phase boundary in the phase diagram of the local interaction and temperature.}
\label{Stoner_TCr}
\end{figure}

\begin{figure}[t]
\includegraphics[width=0.45\textwidth]{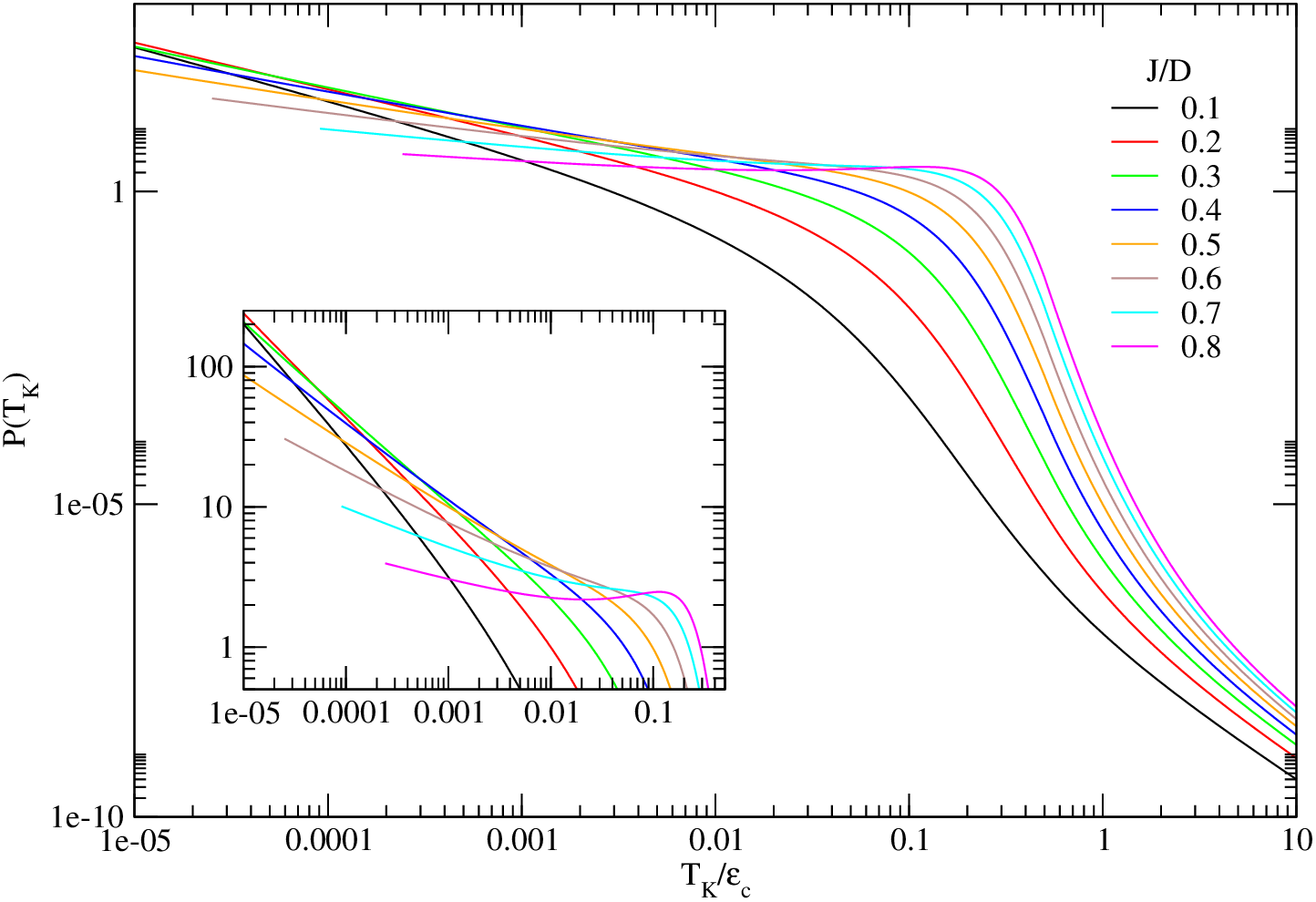}
\caption{A distribution function of the Kondo temperature. The distribution function of the Kondo temperature is given by $P(T_K) = \left|\frac{dT_K}{d \alpha} \right|^{-1} P(\alpha)$ with the mean-field equation (27), where the information of the wave-function multifractality is introduced. An essential point is that it shows a power-law increasing behavior for all Kondo interactions of $J < 0.8 D$, approaching the zero Kondo temperature. This power-law physics results from rare events, meaning that a local magnetic moment remains unscreened, which originates from the local pseudogap region. The inset clarifies this quantum Griffiths physics.}
\label{Kondo_PTK}
\vspace*{1cm}
\includegraphics[width=0.45\textwidth]{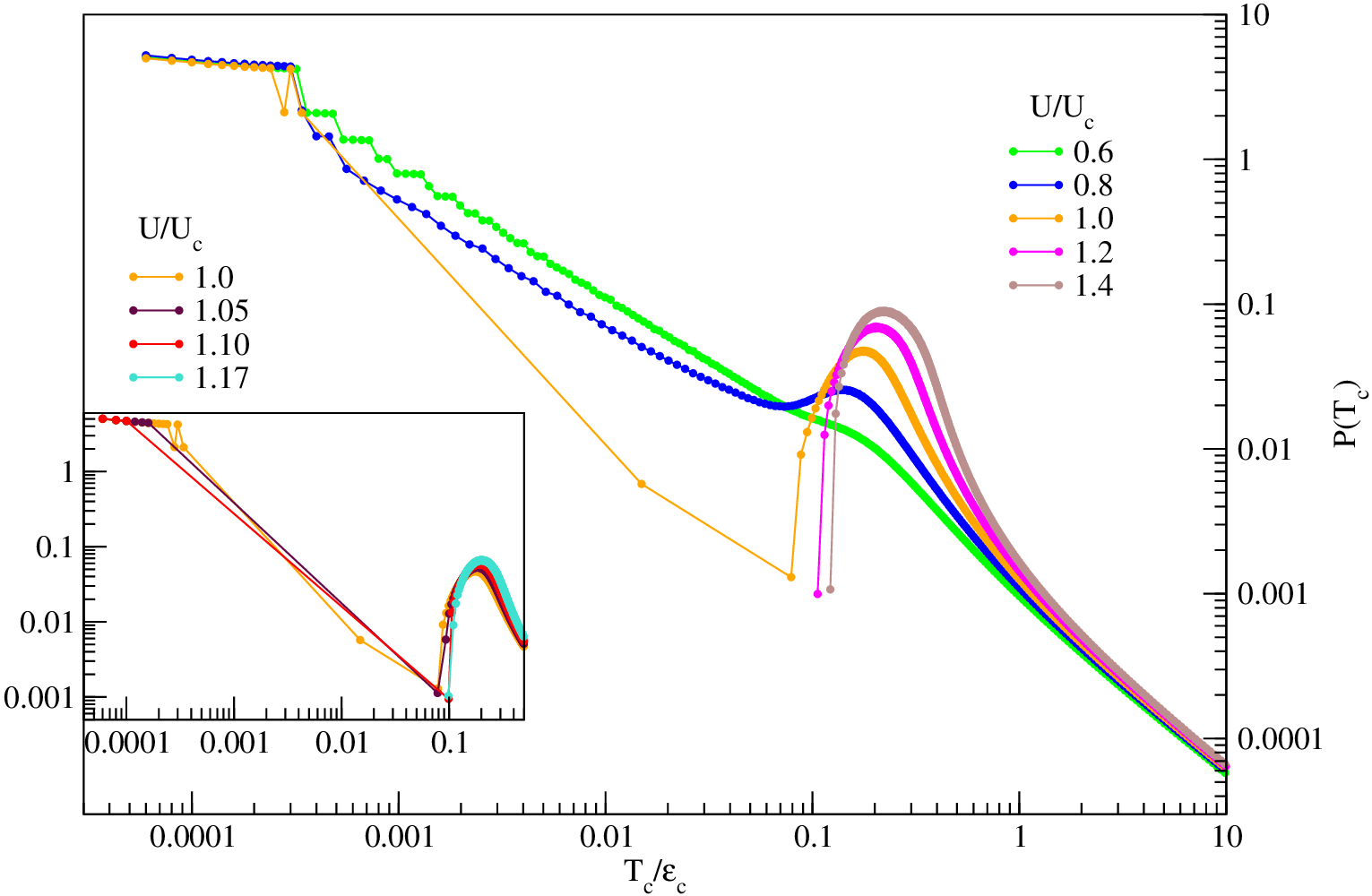}
\caption{A distribution function of the Stoner transition temperature. The distribution function of the local critical temperature is given by $P(T_c) = \left|\frac{dT_c}{d \alpha} \right|^{-1} P(\alpha)$ with the mean-field equation (18), where the information of the wave-function multifractality is introduced. The power-law increasing behavior stops in $1.1 < U/U_{c} < 1.2$, where the distribution function vanishes in $T_{c} \leq 0.1 \varepsilon_{c}$ above $U \sim 1.2 U_{c}$, more clarified in the inset figure. This results from the fact that the critical temperature changes discontinuously in the pseudogap region ($r_{\alpha} > 0$) near $U \sim U_{c}$, giving rise to $\left|\frac{dT_c}{d \alpha} \right| \rightarrow \infty$ as $T_{c} \rightarrow 0$.}
\label{Stoner_PTC}
\end{figure}

The local Kondo temperature decreases to vanish at the critical eigenfunction intensity, $r_\alpha^{c} = J_{K} \rho$, increasing $r_{\alpha}$ in a given Kondo interaction, which corresponds to reducing the local density of states, where $\rho(\varepsilon_{n}) = \rho_{m} \left|\frac{\varepsilon_n}{\varepsilon_c}\right|^{r_\alpha}$. Larger Kondo interactions enhance the Kondo temperature in a given $r_{\alpha}$, i.e., given local density of states. See Fig. \ref{Kondo_TKr}. Essentially the same trend has been observed in the Stoner transition. However, there exists an important different aspect between these two cases: Increasing local interactions, we find that the local critical temperature of the Stoner transition drops down abruptly above a certain positive value of $r_{\alpha}$. In other words, the critical temperature changes discontinuously from a finite lowest critical temperature to the zero critical temperature at a certain positive $r_{\alpha}$ above the critical interaction parameter. See Fig. \ref{Stoner_TCr}. Here, the positive $r_{\alpha}$ means that the local region is in a pseudogap state, where the density of states vanishes in a power-law fashion, approaching the zero energy. This observation leads us to conclude that the local pseudogap region ($r_{\alpha} > 0$) shows an abrupt phase boundary in the Stoner phase diagram of the local interaction and temperature. On the other hand, the local Kondo temperature is well defined to decrease continuously in the pseudogap region until it vanishes. It turns out that the hybridization order parameter decreases to vanish continuously, exhibiting a conventional ``second-order transition" in the phase diagram of the Kondo coupling and temperature with a given $r_{\alpha}$.

The relation between the local critical temperature and the local eigenfunction intensity or the local density of states allows us to translate the log-normal distribution function for the eigenfunction intensity into a power-law distribution function for the local critical temperature. The distribution function of the local Kondo temperature shows a power-law increasing behavior for all Kondo interactions of $J < 0.8 D$, approaching the zero Kondo temperature. See Fig. \ref{Kondo_PTK}. This means that the local magnetic moment remains unscreened, which originates from the local pseudogap region. Mathematically, the high probability of the local moment physics comes from $\left|\frac{dT_K}{d \alpha} \right| \rightarrow 0$ as $r_{\alpha} \rightarrow J_{K} \rho$. As a result, the typical value of the Kondo temperature vanishes identically, where the typical value is defined as a geometric average \bqa && \langle T_{K} \rangle_{typ} \equiv \exp\Bigl\{ \int_{0}^{\infty} d T_{K} P(T_{K}) \ln T_{K} \Bigr\} . \eqa On the other hand, the power-law increasing behavior in the distribution function of the ferromagnetic critical temperature disappears in $1.1 < U/U_{c} < 1.2$, where the distribution function vanishes in $T_{c} \leq 0.1 \varepsilon_{c}$ above $U \sim 1.2 U_{c}$. See Fig. \ref{Stoner_PTC}. This results from the fact that the critical temperature changes discontinuously in the pseudogap region ($r_{\alpha} > 0$) near $U \sim U_{c}$, giving rise to $\left|\frac{dT_c}{d \alpha} \right| \rightarrow \infty$ as $T_{c} \rightarrow 0$ and thus, $P(T_{c} \rightarrow 0) \rightarrow 0$. As a result, the typical value of the ferromagnetic transition temperature vanishes identically in $U < 1.2 U_{c}$ while it becomes finite above this characteristic value of the interaction parameter. The typical transition temperature is expected to change discontinuously.

\begin{figure}[t]
\includegraphics[width=0.48\textwidth]{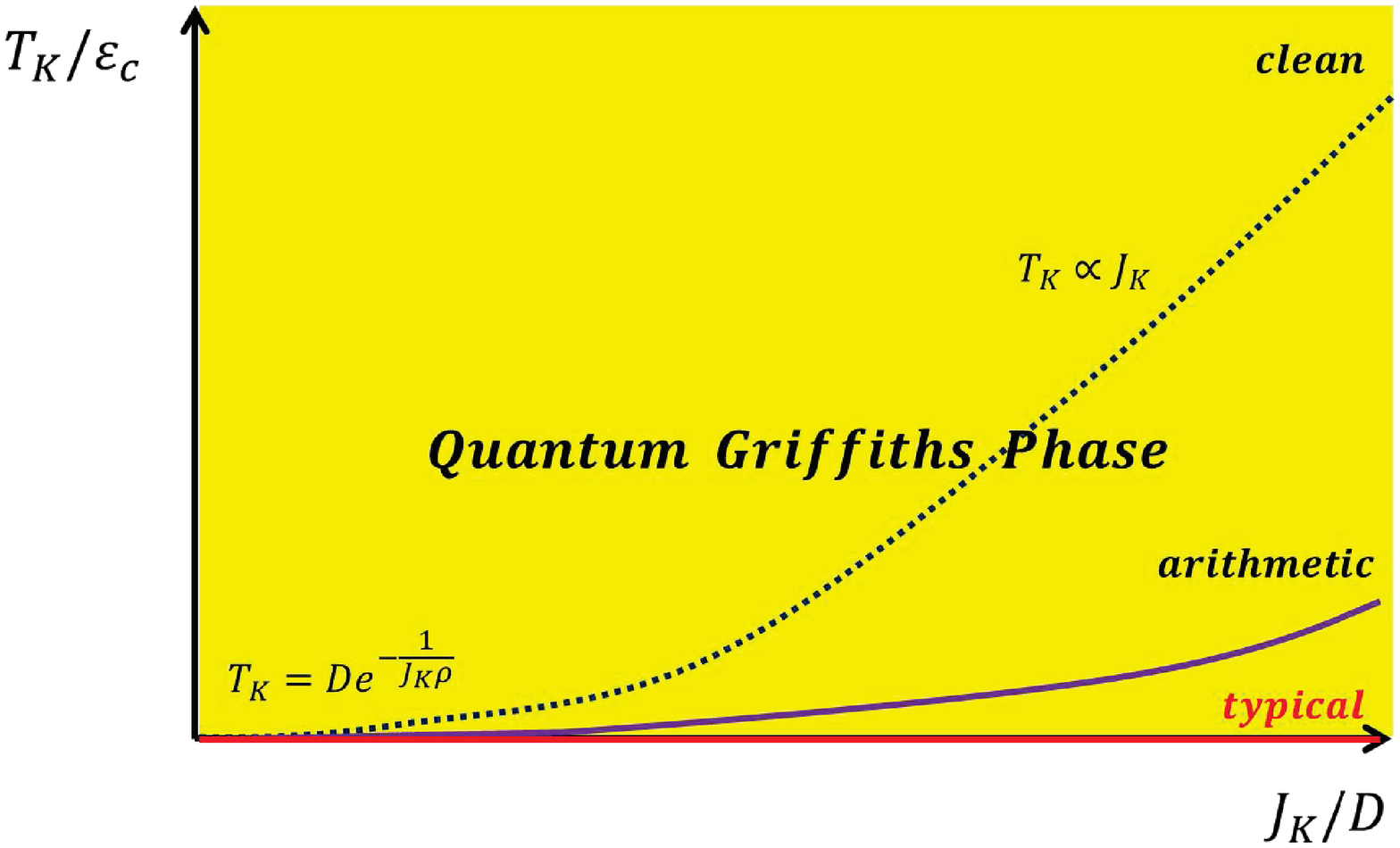}
\caption{A schematic Kondo phase diagram at the Anderson metal-insulator transition. A characteristic feature is that the typical Kondo temperature vanishes due to dominant behaviors of local pseudogap regions, thus governed by decoupled local moment physics. This leads us to conclude that the finite temperature region shows quantum Griffiths phenomena, where the existence of the power-law tail in the distribution function is responsible for quantum Griffiths effects.}
\label{Kondo_phase_diagram}
\vspace*{1cm}
\includegraphics[width=0.48\textwidth]{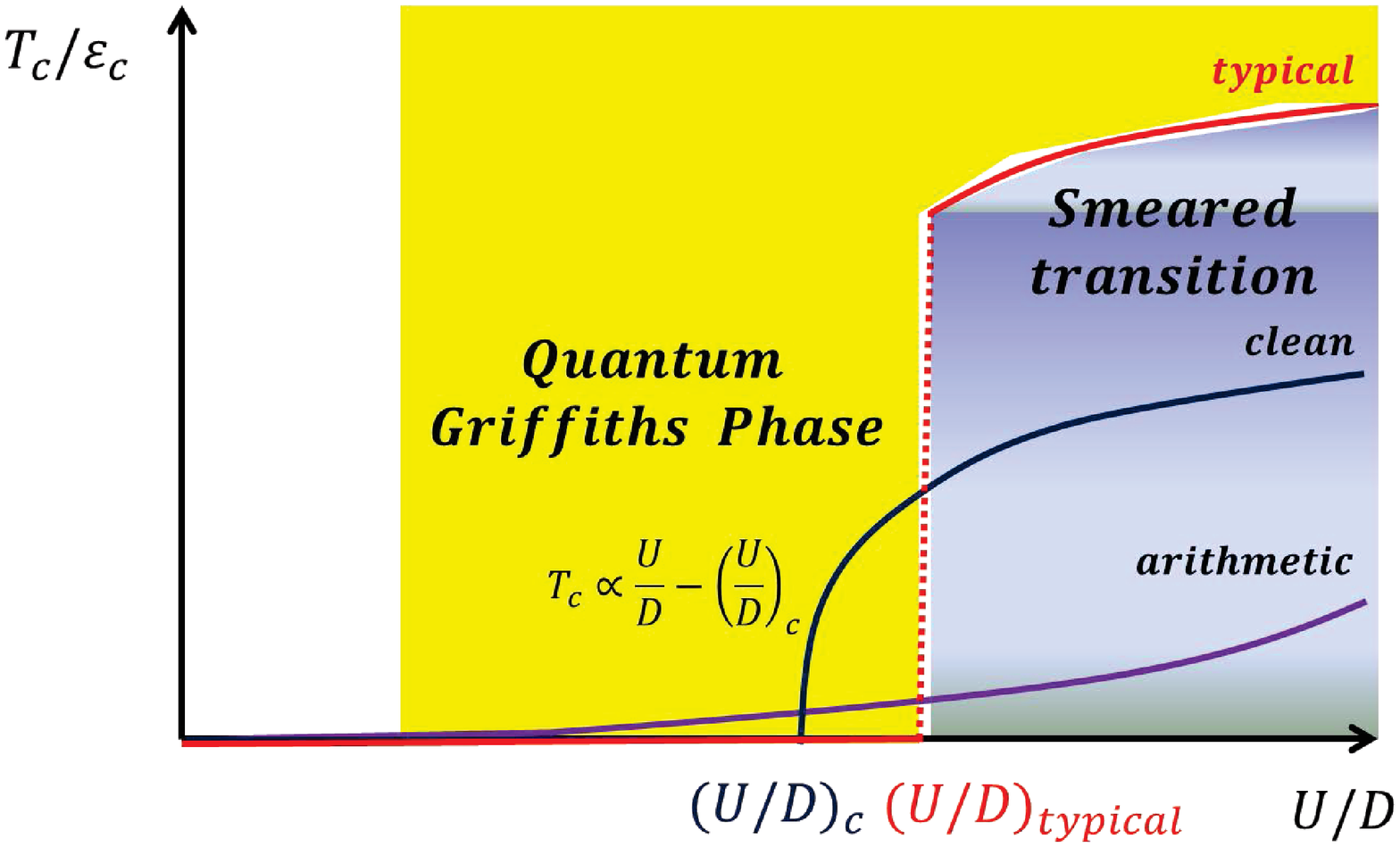}
\caption{A schematic Stoner phase diagram at the Anderson metal-insulator transition. A noticeable feature is that the typical transition temperature increases discontinuously as a function of the interaction parameter, where it vanishes in $U < 1.2 U_{c}$ but it becomes finite above the characteristic interaction parameter. This discontinuous enhancement around the characteristic interaction parameter results from the disappearance of the power-law tail in the distribution function of the critical temperature. As a result, quantum Griffiths effects disappear below the typical transition temperature since the local region is already ferromagnetically ordered. It is natural to suspect that the phase transition across this typical transition temperature is smeared in nature since such a local region is already ferromagnetically ordered and the ordering temperature should be broadened around the typical transition temperature.}
\label{Stoner_phase_diagram}
\end{figure}

These typical local transition temperatures lead us to propose phase diagrams for the Kondo effect and the Stoner transition, respectively, when dynamics of electrons lies at the Anderson metal-insulator transition. See Figs. \ref{Kondo_phase_diagram} and \ref{Stoner_phase_diagram}. The crossover Kondo temperature from a decoupled local moment state to a local Fermi-liquid state is well known in the clean limit,
%
%
given by $T_{K} \sim D e^{- \frac{1}{J_{K} \rho}}$ in the weak-coupling limit and $T_{K} \sim J_{K}$ in the strong Kondo coupling regime \cite{Hewson_Kondo}. This Kondo temperature turns out to be much suppressed at the Anderson transition, measured in the arithmetic average $\langle T_{K} \rangle = \int_{0}^{\infty} d T_{K} P(T_{K}) T_{K}$. Here, we focus on the typical Kondo temperature, most probable and thus, regarded to be a reasonable measure for a phase transition, more correctly, a crossover energy scale. We claim that quantum Griffiths phenomena occur when the measuring temperature is above the typical transition temperature, dominated by physics of rare events. As shown in the above, the typical Kondo temperature turns out to vanish due to dominant behaviors of local pseudogap regions, thus governed by decoupled local moment physics. As a result, we conclude that the finite temperature region of the Kondo effect at the Anderson transition shows quantum Griffiths phenomena, where anomalous power-law physics are expected to appear (Fig. \ref{Kondo_phase_diagram}).

%
%
In the Stoner transition the arithmetically averaged transition temperature is much suppressed as the phase diagram of the Kondo effect, compared with the transition temperature of the clean case. A noticeable feature is that the typical transition temperature shows an abrupt change as a function of the interaction parameter, where it vanishes in $U < 1.2 U_{c}$ but it becomes finite above the characteristic interaction parameter (Fig. \ref{Stoner_phase_diagram}). This discontinuous enhancement of the typical transition temperature around the characteristic interaction parameter results from the disappearance of the power-law tail in the distribution function of the critical temperature. Based on this phase diagram, we propose that quantum Griffiths effects would be observed above the typical transition temperature. If we focus on the low-temperature regime, quantum Griffiths phenomena occur below the characteristic interaction parameter and disappear above it, where the power-law tail of the distribution function is gone. The nature of the ferromagnetically ordered state below the typical transition temperature and above the characteristic interaction parameter is not completely clarified. However, it is natural to suspect that the phase transition across this typical transition temperature is smeared in nature since such a local region is already ferromagnetically ordered and the ordering temperature should be broadened around the typical transition temperature. In this respect we propose the typical transition temperature as a criterion for the appearance of either quantum Griffiths phenomena or smeared phase transitions.

\begin{figure}[t]
\includegraphics[width=0.45\textwidth]{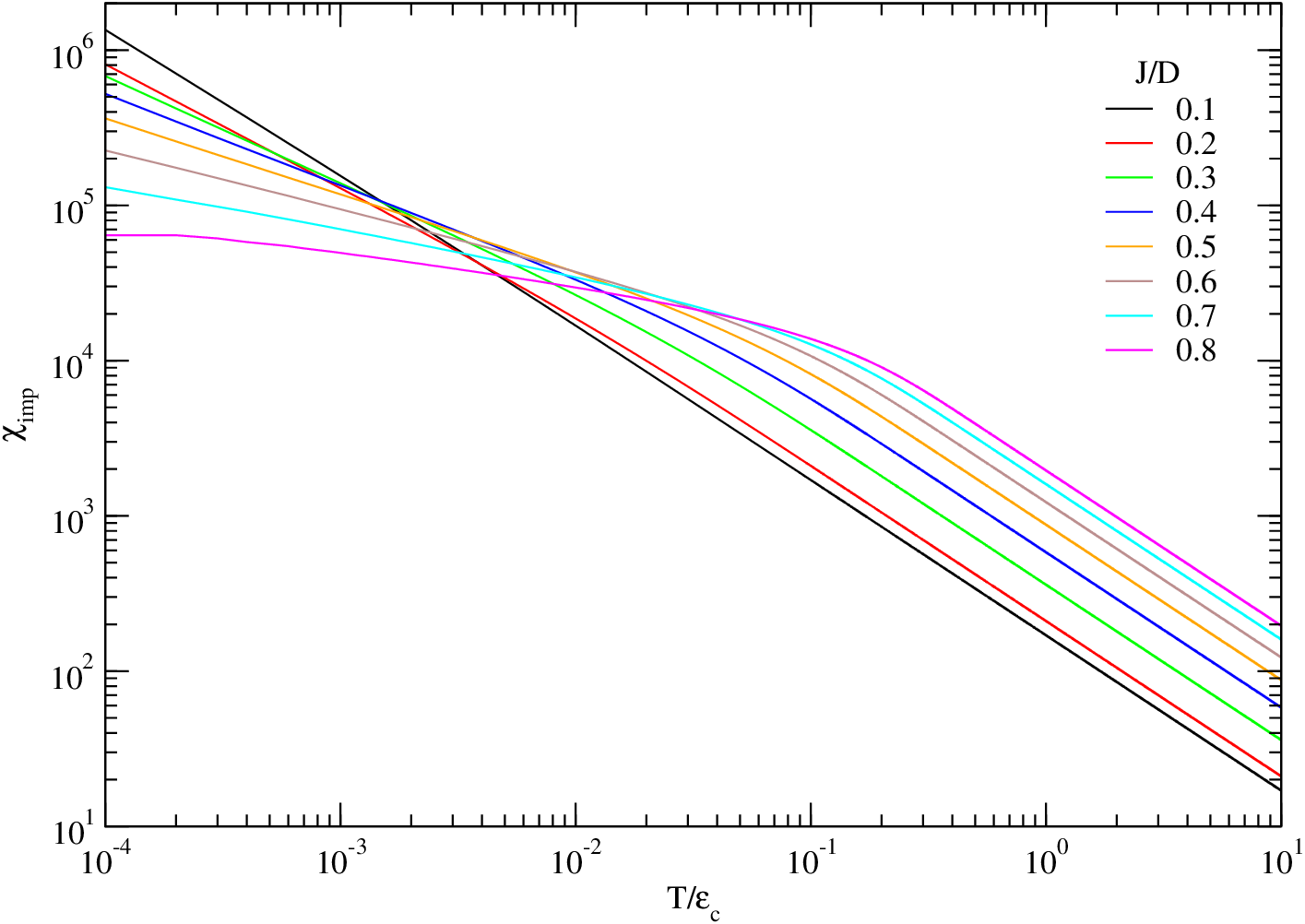}
\caption{Typical impurity spin susceptibility in the Kondo effect. At high temperatures, the first term in Eq. (\ref{Chi_imp}) plays a dominant role in the susceptibility, resulting in the Curie-type behavior. At low temperatures, the second term in Eq. (\ref{Chi_imp}) is a leading contribution, which turns out to be identical to the inverse of the typical Kondo temperature. As a result, the typical impurity spin susceptibility diverges slower than the Curie behavior due to the Kondo effect.}
\label{Ximp}
\vspace*{1cm}
\includegraphics[width=0.45\textwidth]{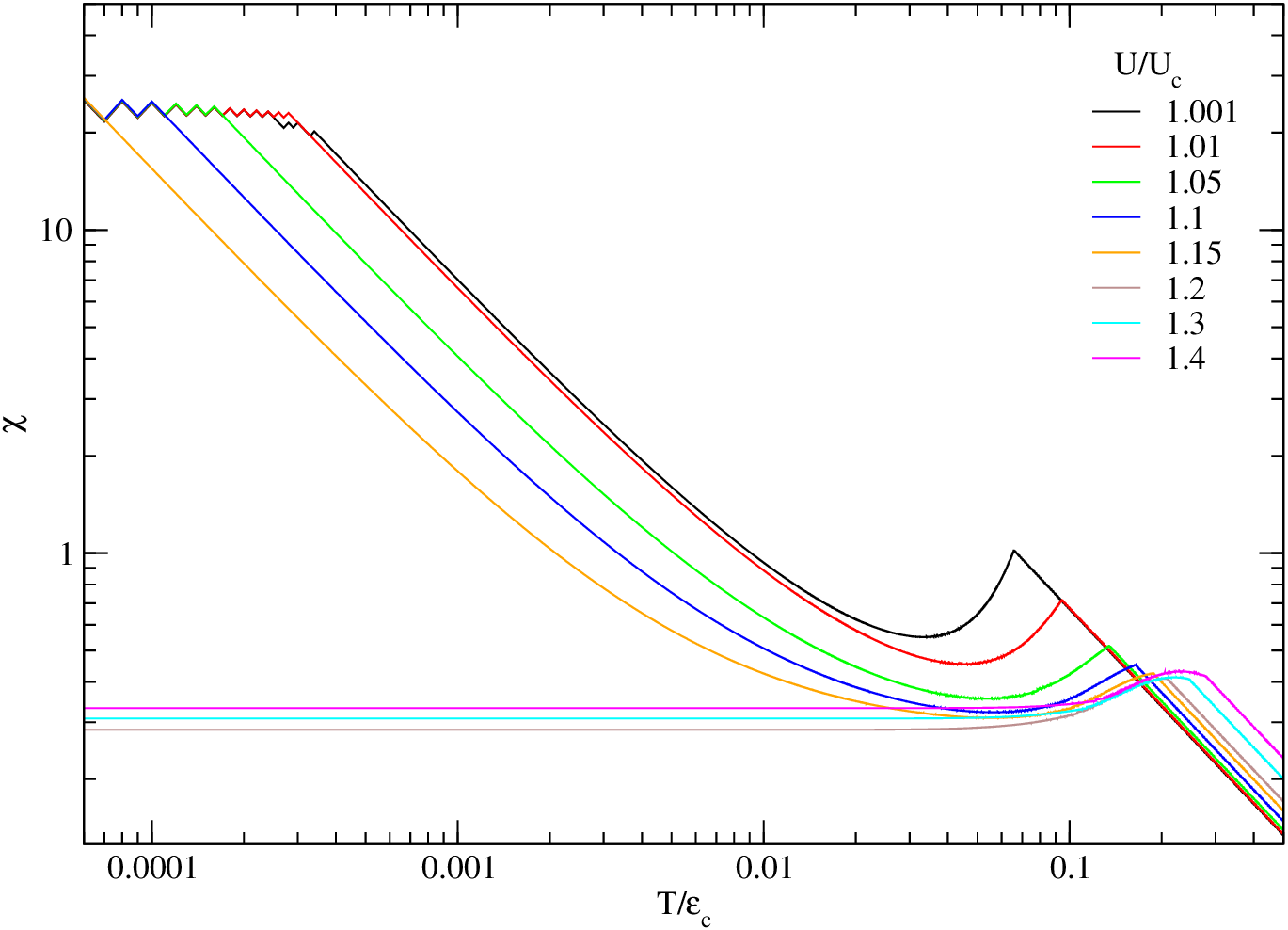}
\caption{Typical spin susceptibility in the Stoner transition. At high temperatures, it results in the Curie-type behavior, given by the first term in Eq. (\ref{Chi_FM}). At low temperatures, the second term in Eq. (\ref{Chi_FM}) plays a central role in the spin susceptibility, identifying the typical local spin susceptibility with the inverse of the typical transition temperature as that of the Kondo effect. Since the typical transition temperature evolves discontinuously from zero to a finite value as a function of the interaction parameter, we find that the divergent behavior of the typical local spin susceptibility above the typical transition temperature and below the characteristic interaction parameter ($\sim 1.2 U_{c}$) disappears to saturate into a finite value, regarded to be the Pauli spin susceptibility, below the typical transition temperature and above the characteristic interaction parameter.}
\label{X_Stoner}
\end{figure}

In order to support that the typical transition temperature is a criterion for the quantum Griffiths and smeared transition phenomena, we evaluate the spin susceptibility in a geometric average, expected to be quite sensitive to the typical transition temperature. We call this quantity the typical spin susceptibility. The typical impurity spin susceptibility can be calculated as follows
%
%
\begin{eqnarray}
 \chi_{imp}^{typ}(T) & = & \exp\Bigl\{ \int_{0}^T dT_K P(T_K) \ln \chi(T > T_{K}) \nn &+& \int_T^{\infty} dT_K P(T_K) \ln \chi(T < T_{K}) \Bigr\} , \label{Chi_imp}
\end{eqnarray}
where $\chi(T < T_{K}) = \frac{C}{T_K}$ and $\chi(T > T_{K}) = \frac{C}{T}$ with a positive numerical constant $C$ associated with the spin quantum number of an impurity. It is straightforward to estimate both high and low temperature limits of the typical impurity spin susceptibility. At high temperatures, the first term plays a dominant role in the susceptibility, resulting in the Curie-type behavior. At low temperatures, the second term is a leading contribution, which turns out to be identical to the inverse of the typical Kondo temperature. As a result, the typical impurity spin susceptibility diverges, approaching zero temperature, slower than the Curie behavior due to the Kondo effect. This estimation is indeed confirmed in Fig. \ref{Ximp}, where the typical impurity spin susceptibility gives a power-law divergent behavior.

The typical local spin susceptibility in the Stoner transition can be evaluated as follows
%
%
\bqa
 \chi_{FM}^{typ}(T) & = & \exp\Bigl\{ \int_{0}^T d T_c P(T_c) \ln \chi(T > T_c) \nn &+& \int_T^{\infty} d T_c P(T_c) \ln \chi(T < T_c) \Bigr\} , \label{Chi_FM}
\eqa
where $\chi(T < T_c) = \frac{\tanh \left[\frac{m(T)}{2 T} \right]}{m(T)}$ and $\chi(T > T_c) = \frac{1}{2 T}$ with the magnetization order parameter of $m(T) = m \sqrt{T_{c} - T}$. At high temperatures, it results in the Curie-type behavior, given by the first term. At low temperatures, the second term plays a central role in the spin susceptibility. Taking the approximation of $\chi(T \ll T_c) = \frac{\tanh \left[\frac{m \sqrt{T_{c} - T}}{2 T} \right]}{m \sqrt{T_{c} - T}} \approx \frac{\tanh \left[\frac{m \sqrt{T_{c}}}{2 T} \right]}{m \sqrt{T_{c}}} \approx \frac{1}{m \sqrt{T_{c}}}$, the typical local spin susceptibility becomes the inverse of the typical transition temperature as that of the Kondo effect. The typical transition temperature evolves discontinuously from zero to a finite value as a function of the interaction parameter. As a result, we conclude that the divergent behavior of the typical local spin susceptibility above the typical transition temperature and below the characteristic interaction parameter ($\sim 1.2 U_{c}$) disappears to saturate into a finite value, regarded to be the Pauli spin susceptibility, below the typical transition temperature and above the characteristic interaction parameter. The crossover behavior from quantum Griffiths phenomena to smeared phase transitions is also seen in the typical local spin susceptibility for the Stoner transition at the Anderson metal-insulator transition. See Fig. \ref{X_Stoner}.

\section{Summary and discussion}

The argument based on the lower critical dimension of rare events serves an intuitive picture for a criterion on quantum Griffiths phenomena vs. smearing phase transitions at zero temperature \cite{Vojta_Review}. It is not clear how to generalize this physical picture toward a finite temperature region. In this study we proposed a criterion, expected to work at finite temperatures. It is not surprising to introduce a typical transition temperature as the ``critical" temperature at the Anderson metal-insulator transition. An essential question is how to calculate the typical transition temperature. An idea was to construct a mean-field theory for the symmetry breaking transition of rare regions, where the wave-function multifractality is introduced to impose the role of Anderson localization in phase transitions. It is true that this local mean-field theory framework does not take into account the information involved with correlations between local regions (correlated inhomogeneity). However, investigating the local critical temperature as a function of the eigenfunction intensity, the local mean-field theory construction can be justified within a granular picture, where correlations between rare events (granules) may not be relevant. See the discussion of section II-A-3 with Ref. \cite{Discussion_Inhomogeneity}. As a result, we could obtain the relation between the critical temperature and the eigenfunction intensity. This allowed us to translate the log-normal distribution function of the eigenfunction intensity into the power-law distribution function of the critical temperature. Then, it was straightforward to calculate the typical value of the transition temperature, resorting to the distribution function of the critical temperature.

We investigated two kinds of ``phase transitions": the Kondo effect vs. the Stoner transition. It turns out that the typical Kondo temperature vanishes for all Kondo interactions in $J_{K} < 0.8 D$, originating from the persistence of the power-law tail up to the zero temperature in the distribution function of the Kondo temperature. Within the local mean-field theory framework, it is clear that this power-law divergent distribution function results from the role of the local pseudogap region, regarded to be a rare event. On the other hand, the typical ferromagnetic transition temperature evolves discontinuously as a function of the local interaction parameter, where it vanishes below the characteristic interaction parameter about $U \sim 1.2 U_{c}$ but it becomes finite abruptly above the interaction parameter. This behavior is also rooted in the fact that the power-law divergent behavior in the distribution function of the Stoner transition temperature disappears to drop down in a certain temperature range near the typical transition temperature above the characteristic interaction parameter. These typical temperatures lead us to propose phase diagrams for the Kondo effect and the Stoner transition, respectively, at the Anderson metal-insulator transition. Since the typical Kondo temperature vanishes identically, we suggested quantum Griffiths effects at finite temperatures in this disordered Kondo system. On the other hand, since the typical Stoner transition temperature is finite above the characteristic interaction parameter, we claimed that quantum Griffiths phenomena above the typical transition temperature disappears replaced by smeared phase transitions due to preexisting ferromagnetic ordering.

In order to support this physical picture, we calculated the typical spin susceptibility, which turns out to be connected to the typical transition temperature quite closely at low temperatures. Indeed, we observed that the typical local spin susceptibility is given by the inverse of the typical transition temperature at low temperatures. The typical impurity spin susceptibility showed its divergent behavior, where the divergent degree is weaker than that of the Curie-type behavior due to the Kondo effect. Such a power-law divergent behavior, observed in the whole Kondo-interaction range, is consistent with quantum Griffiths phenomena. One the other hand, the typical local spin susceptibility diverges at low temperatures below the characteristic interaction parameter, but it becomes saturated into a finite value below the typical transition temperature and above the characteristic interaction parameter, consistent with the behavior of smeared phase transitions.

An important issue, not discussed in the present study, is on the role of repulsive interactions between density fluctuations in both the Kondo effect and the Stoner transition. Such effective interactions may be introduced into this mean-field theory framework via the Hartree-Fock approximation, described by additional mean-field equations and expected to cause the Altshuler-Aronov suppression of the density of states \cite{AAcorrection_inDFL}. As a result, pseudogap physics identified with rare events would promote quantum Griffiths effects more than the present situation without repulsive interactions. It is important to perform the full numerical analysis for the Hartree-Fock theory with ferromagnetic transition, where both self-consistent renormalizations for interactions and disorders and strong spatial fluctuations in dynamics of order parameters can be taken into account.

\section*{Acknowledgement}

This study was supported by the Ministry of Education, Science, and Technology (No. NRF-2015R1C1A1A01051629 and No. 2011-0030046) of the National Research Foundation of Korea (NRF) and by TJ Park Science Fellowship of the POSCO TJ Park Foundation. This work was also supported by the POSTECH Basic Science Research Institute Grant (2015). We would like to appreciate fruitful discussions in the APCTP workshop on Delocalisation Transitions in Disordered Systems in 2015. We thank S. Kettemann for fruitful collaborations and insightful discussions at the initial stage. KS also appreciates enlightening discussions with V. Dobrosavljevic.


\begin{thebibliography}{9}
\bibitem{Alpha} We have $\alpha_{n}(\bm{r}) = - \frac{\ln (L/\xi_n)^d |\Psi_{n}(\bm{r})|^{2}}{\ln \xi_n}$ in the metallic regime $\varepsilon_n > \varepsilon_{m}$, where $\xi_n$ denotes the correlation length. In the insulating regime $\varepsilon_n < \varepsilon_{m}$, we get $\alpha_{n}(\bm{r}) = - \frac{\ln  |\Psi_{n}(\bm{r})|^{2}}{\ln L_{c n}}$, where $L_{c n}$ is the localization length \cite{PTK2}. In both cases the distribution function is Gaussian. Since interactions give rise to correlations between eigenfunctions in different energies, all of these expressions are necessary to be taken into account.
\bibitem{Energy_Level_Comment} The single-particle energy spectrum is assumed to be picket-fence-like (equally spaced levels) and non-fluctuating. It is well-known that eigenenergy fluctuations play a minor role in the statistics of the Kondo temperatures \cite{PTK2}.
\bibitem{Discussion_Inhomogeneity} Expanding the Fermi-Dirac distribution function up to the first order of local magnetization in Eq. (\ref{Self_Consistent_Equation}), we obtain \bqa && {\Phi}(\bm{r}) = \frac{U \rho_{m}}{6} \int_{-\varepsilon_{c}}^{\varepsilon_{c}} d \varepsilon_{n} \frac{\int d^{d} \bm{r'} \Bigl\langle |\Psi_{n}(\bm{r})|^{2} |\Psi_{n}(\bm{r'})|^{2} \Bigr\rangle {\Phi}(\bm{r'})}{\cosh^{2}\Bigl(\frac{\varepsilon_{n}}{2T_{c}(\bm{r})}\Bigr)} , \nonumber  \eqa where $\Bigl\langle ... \Bigr\rangle$ is given by Eq. (\ref{Average_Multifractality}). Comparing this equation with Eq. (\ref{Local_Tc}), we conclude that the nonlocality becomes irrelevant if \bqa && \int d^{d} \bm{r'} \Bigl\langle |\Psi_{n}(\bm{r})|^{2} |\Psi_{n}(\bm{r'})|^{2} \Bigr\rangle {\Phi}(\bm{r'}) \approx \Bigl|\frac{\varepsilon_{n}}{\varepsilon_{c}}\Bigr|^{2 r_{\alpha}} {\Phi}(\bm{r}) \nonumber \eqa is satisfied within a certain length scale.
\end{thebibliography}
\end{document}